\documentclass{Interspeech}

\interspeechcameraready

\title{Exploring the Potential of Large Multimodal Models as Effective Alternatives for Pronunciation Assessment}

\author[affiliation={1}]{Ke}{Wang}
\author[affiliation={1}]{Lei}{He}
\author[affiliation={1}]{Kun}{Liu}
\author[affiliation={1}]{Yan}{Deng}
\author[affiliation={1}]{Wenning}{Wei}
\author[affiliation={1}]{Sheng}{Zhao}
\affiliation{Microsoft}{Beijing}{China}
\email{\{wake,helei,v-liukun,yaden,wennwei,szhao\}@microsoft.com}

\usepackage{comment}

\begin{document}

\maketitle

\begin{abstract}

    Large Multimodal Models (LMMs) have demonstrated exceptional performance across a wide range of domains. This paper explores their potential in pronunciation assessment tasks, with a particular focus on evaluating the capabilities of the Generative Pre-trained Transformer (GPT) model, specifically GPT-4o. Our study investigates its ability to process speech and audio for pronunciation assessment across multiple levels of granularity and dimensions, with an emphasis on feedback generation and scoring. For our experiments, we use the publicly available Speechocean762 dataset. The evaluation focuses on two key aspects: multi-level scoring and the practicality of the generated feedback. Scoring results are compared against the manual scores provided in the Speechocean762 dataset, while feedback quality is assessed using Large Language Models (LLMs). The findings highlight the effectiveness of integrating LMMs with traditional methods for pronunciation assessment, offering insights into the model's strengths and identifying areas for further improvement.
\end{abstract}

\section{Introduction}

Large Language Models (LLMs) have recently captivated significant interest, showcasing impressive results in diverse fields, including understanding and generating human-like text, creating immersive visual content, and interpreting complex auditory data\cite{devlin2018bert,ramesh2021valle,openai2024sora,chen2022wavlm,wang2023valle}. Among LLMs, the OpenAI-developed Generative Pre-trained Transformer (GPT) \cite{radford2018gpt,radford2019gpt2,openai2020gpt3,openai2024gpt4} is one of the most influential models, along with its variants \cite{meta2024llama3,google024gemini,mistral2024pixtral12b,alibaba2024qwen25,deepseek2024deepseekv3}. The advent of GPT-4o \cite{openai2024gpt4o} and Gemini 2.0 \cite{google2024gemini20}, which amalgamate generative and multimodal Artificial Intelligence (AI), signifies the advancement of Large Multimodal Models (LMMs) towards facilitating more natural human-computer interaction with minimal latency and an enhanced capacity to handle more intricate tasks.

Education remains a pivotal component in human societal development and advancement. However, traditional education grapples with issues such as individual student disparities and inadequate allocation of educational resources. Additionally, assessing teaching effectiveness continues to pose a significant challenge. As a result, the integration of LLMs in the educational sector \cite{gan2023llms} is gaining momentum to mitigate these problems.

In the realm of Computer-Assisted Language Learning (CALL), Automated Essay Scoring (AES) and Pronunciation Assessment (PA) are two primary components. AES, employed for written language evaluation, predominantly concentrates on content scoring, encapsulating the assessment of writing skills such as grammar, vocabulary, coherence, and organization. Conversely, PA, utilized for spoken language evaluation, emphasizes speaking skills, assessing elements like pronunciation accuracy, fluency, and prosody. The implementation of LLMs in AES is direct and has yielded impressive results in scoring and feedback generation \cite{han2023llm,xiao2024human,stahl2024exploring,banno2024gpt4,kim2024is}. This is attributed to their robust performance in language processing and their ability to learn in context.

In recent years, LLMs have ventured into the realm of PA tasks, despite models like GPT lacking inherent speech and audio capabilities. For example, ChatGPT \cite{openai2022chatgpt}, leveraging zero-shot and few-shot learning, was employed to probe into the potential of LLMs in speech phrase break assessment \cite{wang23assessing}. Given the lack of speech and audio processing abilities, the speech snippets were preprocessed into text using speech-to-text force alignment. The findings revealed that ChatGPT (text-chatdavinci-002) struggled to effectively address breaks between semantic groups and consistently demonstrated inconsistent performance when attempting to correct improper breaks. Inspired by the significant impact of LLMs on text-related scoring tasks, a PA model that integrates LMMs was proposed to tackle sentence-level accuracy and fluency scoring \cite{fu2024pronunciation}. In this model, speech was initially mapped into contextual features using a pre-trained acoustic encoder. A modality adapter was subsequently utilized to merge the acoustic features with the text features. Lastly, the multimodal features were inputted into the LMMs for pronunciation scoring. The results showcased that the multimodal LMMs achieved competitive results compared to traditional models concerning sentence-level accuracy and fluency scores.

In this paper, we propose to utilize LMMs for PA tasks using zero-shot learning. To the best of our knowledge, this is the first study to delve into the objective evaluation of LMMs for pronunciation assessment across multiple granularities and dimensions, and to offer comprehensive feedback. Due to the constraints of the GPT-4o real-time API, which solely accepts text prompts, we refrain from exploring one- and few-shot learning in this study.

\section{Related work}

PA is a vital element of CALL systems. It entails evaluating language learners' pronunciation quality, including aspects such as accuracy, fluency, and prosody, in order to provide feedback and guide their progression. Existing research on PA can be divided into two categories: align-based methods \cite{witt2000phone,wang2012improved,hu2015improved,mao2019nn,gong2022transformer,chao2023hierarchical} and align-free methods \cite{fu2024pronunciation,kim2022automatic,liu2023asr,liang2023end}.

Align-based methods require an Automatic Speech Recognition (ASR) model to perform force alignment between the audio and the corresponding reference text. Goodness of Pronunciation (GOP) \cite{witt2000phone} and its variants \cite{wang2012improved,hu2015improved} are dominant techniques in this area. After obtaining the aligned features, a scoring network is built to map these features to pronunciation scores.

Conversely, align-free methods directly employ the acoustic and text-related linguistic features to predict pronunciation scores with an end-to-end network. This approach significantly simplifies the feature extraction process and eliminates the influence of force alignment on scoring. Previous studies on align-free methods have demonstrated that they can achieve competitive results in comparison to align-based methods in terms of sentence-level accuracy \cite{fu2024pronunciation}, fluency \cite{fu2024pronunciation,kim2022automatic,liu2023asr}, and prosody \cite{kim2022automatic}. However, align-free methods have not been exhaustively investigated for smaller granularities, such as phoneme- and word-level scoring \footnote{We have communicated with the authors of the \cite{liang2023end} and discovered some errors in their calculation of the final results.}.

Recently, with the advent of GPT-4o \cite{openai2024gpt4o}, which can accept input in any combination of text, audio, image, and video, and generate output in any combination of text, audio, and image, its multimodal capabilities make it suitable for multimodal tasks such as align-free PA. Additionally, there is even a publicly available demo entitled ``Point and Learn Spanish"\footnote{https://vimeo.com/945587424}, which showcases GPT-4o's capabilities for language learning on OpenAI's official website. Although it has displayed impressive subjective results, a detailed objective evaluation of the model, to the best of our knowledge, has yet not been explored. In this paper, we aim at investigating the objective evaluation of LMMs, specifically selecting GPT-4o, for PA at multiple granularities, including phoneme-, word-, and sentence-level scoring. These experiments provide insights into the model's strengths and weaknesses and guide future research directions. To summarize, the contributions of this paper are as follows:

\begin{itemize}[leftmargin=2em]
  \item We explore the objective evaluation of LMMs for pronunciation assessment at multiple granularities and across various dimensions.
  \item The evaluation not only illuminates the potential of GPT-4o in enhancing language learning tools but also identifies areas where traditional models may still hold an advantage.
  \item We propose to integrate the pronunciation assessment service with GPT-4o to enhance the quality of feedback and provide more accurate scores on both small and large granularities.
  \item The findings of this study provide valuable insights that can guide future research on pronunciation assessment tasks using LMMs.
  \end{itemize}

\section{Approach}

We propose to evaluate the GPT-4o model for pronunciation assessment at multiple granularities and provide feedback in an alignment-free mode. By comparing the performance of GPT-4o with traditional models, we aim at exploring the following aspects: 1) the efficacy of LMMs, specifically GPT-4o, in pronunciation assessment tasks; 2) the ability of GPT-4o to provide comprehensive feedback that pinpoints areas for improvement for language learners; 3) the strengths and constraints of LMMs in the context of pronunciation assessment.

\subsection{Prompts}

The prompts for the GPT-4o model are meticulously crafted to assess its capabilities in PA tasks. These prompts are divided into five categories: phoneme-level, word-level, sentence-level, multigranularity-level, and feedback generation. The phoneme-level prompts aim at evaluating the accuracy of phoneme pronunciation, while the word-level prompts aim at assessing the accuracy of word pronunciation and the stress of the word. The sentence-level prompts are designed to evaluate the accuracy, fluency, prosody, and completeness of sentence pronunciation. The multigranularity-level prompts strive to evaluate all aspects of the sentence, inclusive of feedback generation. The feedback generation prompts are devised to offer guidance to language learners on enhancing their pronunciation. These prompts are designed to encompass a wide array of phonemes, words, sentences, and feedback scenarios to ensure the evaluation is comprehensive. All prompts are exhibited in Appendix \ref{app:prompts} and the scoring rubric is based on the Speechocean762 \cite{zhang2021speechocean762} manual scoring metrics. The results are returned in JSON format to facilitate subsequent analysis.

Due to the constraints of the GPT-4 real-time API, our study must focus solely on zero-shot prompts \footnote{We also experimented with a one-shot prompt by integrating the human label and audio in the first turn of the dialog and then evaluating the target audio. However, the correlation between predicted scores and human-labeled scores did not show considerable improvement.}. To enhance performance on multi-step reasoning tasks, such as providing feedback based on multigranularity scores, we adopt the Zero-shot-CoT \cite{kojima2023largelanguagemodelszeroshot}. This involves incorporating the phrase ``Let's think step by step," which elicits Chains of Thought (CoT) from LLMs across a variety of reasoning tasks.

\subsection{Feedback assessment}

Utilizing LLMs to evaluate the quality of generated text has demonstrated consistency with human expert annotations in specific free-text generation tasks \cite{chiang2023large}. As there are no existing automatic metrics for assessing the quality of generated feedback, we employ GPT-4, GPT-4o Mini, and Phi-4 \cite{microsoft2024phi4technicalreport} to gauge this quality. We direct these models to assign an overall helpfulness score ranging from 0 (not helpful) to 10 (very helpful) for each piece of generated feedback. This scoring is based on human-labeled scores at the phoneme, word, and sentence levels. Additionally, a score is provided for the correlation between model and human scores. The prompt and core code used for this evaluation are detailed in Table \ref{tab:prompts} and Appendix \ref{app:feedback}.

Our evaluation concentrates on the helpfulness and correlation of the feedback generated by GPT-4o. The helpfulness score signifies how beneficial the feedback is for language learners, which we consider the most critical aspect of feedback quality. Meanwhile, the correlation score measures the degree to which the feedback aligns with human-labeled scores.

\section{Experiments}

\begin{table*}[th]
  \caption{Comparing the performance of pronunciation assessment on multiple granularities and aspects between GPT-4o and baseline models using the Speechocean762 test dataset.}
  \label{tab:gpt_pcc}
  \centering
  \resizebox{\linewidth}{!}{
  \begin{tabular}{c|cc|ccc|ccccc|cc}
    \bottomrule
    \multirow{2}{*}{Model} & \multicolumn{2}{c|}{Phoneme Score} & \multicolumn{3}{c|}{Word Score (PCC)} & \multicolumn{5}{c|}{Utterance Score (PCC)} & \multirow{2}{*}{\#Unscored} & \multirow{2}{*}{\makecell{Unscored \\ Rate (\%)}} \\
    \cline{2-11}
    & RMSE & PCC & Accuracy & Stress & Total & Accuracy & Fluency & Prosody & Completeness & Total \\
    \hline
    GOPT \cite{gong2022transformer} & 0.292 & 0.612 & 0.533 & 0.291 & 0.549 & 0.714 & 0.753 & 0.760 & 0.155 & 0.742 & 0 & 0.00 \\
    3MH \cite{chao2023hierarchical} & 0.266 & 0.693 & 0.682 & 0.361 & 0.694 & 0.782 & 0.843 & 0.836 & 0.374 & 0.811 & 0 & 0.00 \\
    LMMPA \cite{fu2024pronunciaiton} & - & - & - & - & - & 0.713 & 0.777 & - & - & - & 0 & 0.00 \\
    \hline
    GPT-multi & 0.950 & 0.241 & 0.216 & 0.150 & 0.239 & 0.459 & 0.418 & 0.406 & 0.116 & 0.445 & 1,038 & 41.52 \\
    GPT-multi-2 & 0.879  & 0.249 & 0.260 & 0.252 & 0.277 & 0.432 & 0.430 & 0.415 & 0.133 & 0.458 & 1,194 & 47.76 \\
    GPT-phoneme & 0.575 & 0.211 & - & - & - & - & - & - & - & - & 22,797 & 48.13 \\
    GPT-word & - & - & 0.271 & 0.158 & 0.277 & - & - & - & - & - & 7,662 & 47.99 \\
    GPT-sentence & - & - & - & - & - & 0.471 & 0.459 & 0.443 & 0.259 & 0.502 & 1,096 & 43.84 \\
    \hline
    Azure PA & - & - & 0.623 & - & - & 0.700 & 0.715 & 0.842 & 0.258 & 0.782 & 0 & 0.00 \\
    \toprule
  \end{tabular}
  }
\end{table*}

\begin{table*}[th]
  \caption{The SCC of multiple granularity and aspect pronunciation assessment tasks with GPT-4o on the Speechocean762 test dataset.}
  \label{tab:gpt_scc}
  \centering
  \resizebox{\linewidth}{!}{
  \begin{tabular}{c|cc|ccc|ccccc|cc}
    \bottomrule
    \multirow{2}{*}{Model} & \multicolumn{2}{c|}{Phoneme Score} & \multicolumn{3}{c|}{Word Score (SCC)} & \multicolumn{5}{c|}{Utterance Score (SCC)} & \multirow{2}{*}{\#Unscored} & \multirow{2}{*}{\makecell{Unscored \\ Rate (\%)}}\\
    \cline{2-11}
    & RMSE & SCC & Accuracy & Stress & Total & Accuracy & Fluency & Prosody & Completeness & Total \\
    \hline
    GPT-multi & 0.950 & 0.167 & 0.197 & 0.116 & 0.197 & 0.407 & 0.363 & 0.362 & 0.088 & 0.417 & 1,038 & 41.52 \\
    GPT-multi-2 & 0.879 & 0.162 & 0.216 & 0.193 & 0.221 & 0.392 & 0.358 & 0.338 & 0.127 & 0.379 & 1,194 & 47.76 \\
    GPT-phoneme & 0.575 & 0.123 & - & - & - & - & - & - & - & - & 22,797 & 48.13 \\
    GPT-word & - & - & 0.197 & 0.163 & 0.205 & - & - & - & - & - & 7,662 & 47.99 \\
    GPT-sentence & - & - & - & - & - & 0.445 & 0.426 & 0.417 & 0.136 & 0.463 & 1,096 & 43.84 \\
    \hline
    Azure PA & - & - & 0.465 & - & - & 0.677 & 0.617 & 0.783 & 0.143 & 0.749 & 0 & 0.00 \\
    \toprule
  \end{tabular}
  }
\end{table*}

\begin{table*}[th]
  \caption{Feedback evaluation results for GPT-4, GPT-4o mini, and Phi-4 averaged helpfulness and correlation scores on Speechocean762 test dataset.}
  \label{tab:feedback}
  \centering
  \resizebox{\linewidth}{!}{
  \begin{tabular}{c|c|ccc|ccc|ccc|cc}
    \bottomrule
    \multirow{2}{*}{Granularity} & \multirow{2}{*}{\#Feedback} & \multicolumn{3}{c|}{GPT-4} & \multicolumn{3}{c|}{GPT-4o mini} & \multicolumn{3}{c|}{Phi-4} & \multicolumn{2}{c}{Average} \\
    \cline{3-13}
    & & Help. & Corr. & \#Unscored & Help. & Corr. & \#Unscored & Help. & Corr. & \#Unscored & Helpful. & Corr. \\
    \hline
    Multi & 1462 & 7.10 & 7.25 & 0 & 6.38 & 5.97 & 277 & 5.36 & 5.53 & 4 & 6.28 & 6.25 \\
    Multi-2 & 1306 & 7.31 & 7.32 & 0 & 6.59 & 6.18 & 9 & 5.76 & 5.95 & 6 & 6.55 & 6.48 \\
    Single & \textbf{1705} & 7.51 & 7.19 & 0 & 7.07 & 6.37 & 7 & 6.27 & 6.62 & 0 & 6.95 & 6.73 \\
    Single + Azure PA & 1202 & \textbf{8.10} & \textbf{7.73} & 0 & \textbf{7.74} & \textbf{8.33} & 145 & \textbf{7.24} & \textbf{7.89} & 0 & \textbf{7.69} & \textbf{7.98} \\
    \toprule
  \end{tabular}
  }
\end{table*}

The experiment aims at assessing the effectiveness of multigranularity scoring and the usefulness of the generated feedback. These experiments are performed using English speech clip data from the Speechocean762 dataset \cite{zhang2021speechocean762}. The results are measured using the scoring metrics provided by Speechocean762. In the absence of ground-truth pronunciation feedback, we employ LLMs to automatically evaluate the quality of the feedback. This methodology aligns with the approach used in the research conducted by Stahl et al \cite{stahl2024exploring}. It is essential to note that in our experiments, the term ``GPT-4o'' refers to the ``gpt-4o-realtime-preview'' API with the version dated 2024-12-17.

\subsection{Data}

The Speechocean762 dataset, a freely accessible, open-source resource, is specifically curated for pronunciation assessment. It comprises 5,000 English utterances from 250 English learners who are native Chinese speakers. Each learner contributes 20 recorded sentences. The dataset maintains a balanced representation of speakers in terms of gender, age, and English proficiency. The training and test sets are randomly split, with each set containing 2,500 utterances from 125 speakers.

Speechocean762 offers an extensive range of labeling information. At the sentence level, it provides scores for accuracy, fluency, prosody, completeness, and overall performance for each utterance. At the word level, it gives scores for accuracy, stress, and overall performance for every word. Additionally, an accuracy score, ranging from 0 to 2, is assigned to each phoneme; for simplicity, we have converted this to a linear range of 0 to 10. Each score is annotated by five language experts, with the final averaged score serving as the ground truth. In this study, we limit our evaluation to the test set only.

\subsection{Evaluation metrics}

To evaluate scoring performance, we employ the Pearson Correlation Coefficient (PCC), Spearman's Rank Correlation Coefficient (SCC), and Root Mean Square Error (RMSE) as assessment metrics. Both PCC and SCC are used to determine the correlation between the predicted and human-annotated scores, while RMSE quantifies the absolute difference between these two sets of scores. Although Mean Squared Error (MSE) has been a prevalent metric in past studies \cite{zhang2021speechocean762,gong2022transformer,chao2023hierarchical}, we opt for RMSE as it provides a more interpretable scale, eliminating the need to compare MSE values to the third decimal place.

Previous research \cite{zhang2021speechocean762, gong2022transformer,chao2023hierarchical} commonly utilized PCC as an evaluation metric, which is most fitting for continuous data with a linear relationship. However, SCC is more appropriate for ordinal data when the relationship between variables is monotonic but not necessarily linear. In pronunciation assessment tasks, human scores may vary over time, but they consistently exhibit a natural order. Inspired by this characteristic, Mao et al. \cite{mao2019nn} proposed using ordinal regression to evaluate fluency. Therefore, we also employ SCC to assess performance, taking into account the ranking relationship.

Regarding feedback evaluation, our objective is to measure the usefulness of the feedback generated by the LMMs and its correlation with multigranularity human-annotated scores. We accept the scores predicted by the LLMs as a measure of feedback quality. A score closer to 10 suggests that the feedback is more beneficial, and the correlation is stronger.

\subsection{Multigranularity scoring}

GPT-4o has the ability to assess speech pronunciation by assigning scores across various dimensions such as accuracy, fluency, prosody, and completeness of the spoken audio. It can also provide comprehensive feedback, aiding language learners in practicing and improving their pronunciation skills. By leveraging the advanced functionalities of GPT-4o, it becomes possible to evaluate pronunciation across multiple granularities and dimensions simultaneously. In our assessment of GPT-4o, we excluded utterances that resulted in empty responses, and manually corrected responses with incorrect formats, as indicated in the subsequent tables. For comparison purposes, we also conducted Azure Pronunciation Assessment (Azure PA). However, due to the different phonetic sets used by Azure PA and the Speechocean762 dataset, we have not reported the phoneme-level results of Azure PA.

When comparing the PCC results of GPT-4o, referred to as ``GPT-multi" in Table \ref{tab:gpt_pcc}, with those of the baseline models, we observe a significant performance gap across all granularities and dimensions, indicating substantial room for improving GPT-4o's pronunciation assessment capabilities. In the first run of the multigranularity scoring task, 1,038 utterances failed to produce complete results. Out of these, 1,002 utterances were processed with a correct JSON format, while an additional 460 utterances were returned in an invalid JSON format but contained all scoring items and feedback. This resulted in a service failure rate of 41.52\% when disregarding the format issue. These findings suggest that GPT-4o may not be as proficient as baseline models in accurately predicting pronunciation scores without domain-specific data fine-tuning, presenting challenges in achieving a higher success rate in pronunciation assessments.

In our experiments, the minimum temperature setting for the GPT-4o real-time API was set to 0.6. To assess consistency, we performed the multigranularity scoring task twice. The results of these trials are presented as ``GPT-multi-2" in Table \ref{tab:gpt_pcc}. Analyzing these outcomes, we draw similar conclusions to the first run and observe differences in the scores between the two iterations, particularly for the word-level stress score, indicating inconsistent assessment results. In the second run, 1,194 utterances failed to yield complete assessment results, and 688 of these were also unsuccessful in the first run. Additionally, for the benefit of future research efforts, we have included the SCC results in Table \ref{tab:gpt_scc}, where the findings are denoted as ``GPT-multi" and ``GPT-multi-2".

\subsection{Single granularity scoring}

GPT-4o can assess pronunciation across multiple granularities and dimensions, but its performance falls short compared to conventional models. To investigate its potential, we evaluate its single-granularity performance, as shown in Tables \ref{tab:gpt_pcc} and \ref{tab:gpt_scc}, labeled as ``GPT-phoneme", ``GPT-word", and ``GPT-sentence", respectively. When comparing the single granularity scoring outcomes with the multigranularity scoring results, it is observed that GPT-4o performs better at the sentence level, a higher level of granularity, but shows some regression at lower levels such as the phoneme level. The rate of unscored outcomes in the single granularity scoring task is comparable to that in the multigranularity scoring task (43.84\% vs. 41.52\% vs. 47.76\%). This suggests that GPT-4o has a high failure rate (over 40\%) in predicting pronunciation scores at both single and multiple granularities.

\subsection{Feedback evaluation}

Table \ref{tab:feedback} outlines the results of the automatic feedback evaluation. In this experiment, feedback was generated using GPT-4o with zero-shot learning, and the helpfulness and correlation scores were measured using GPT-4, GPT-4o mini, and Phi-4. These scores were evaluated across three tasks: the multi-granularity feedback task, the single feedback generation task, and the ``single feedback + Azure Pronunciation Assessment" task, as denoted ``single + Azure PA" in Table \ref{tab:feedback}. For the latter task, sentence- and word-level scores were first generated using Azure PA. These scores were then integrated into a prompt, as shown in Table \ref{app:prompts}, and used with GPT-4o to produce feedback.

The average helpfulness and correlation scores of GPT-4o consistently surpass 6 for each task. When comparing the results of multiple and single granularity, we find that the single granularity feedback generation task yields better results than the multiple tasks, particularly with the lowest failure rate (1,705 feedbacks were generated in the 2,500 evaluating samples) to generate the feedback. Moreover, combining the Azure PA results with GPT-4o could yield better results both on helpfulness and correlation. These results suggest that the feedback generated by GPT-4o is useful and correlates well with human-annotated scores. This observation aligns with our subjective impressions that GPT-4o can generate reasonable feedback and offer plausible guidance to language learners, yet it still has some limitations in the quality of feedback at smaller granularities, such as at the phoneme and word levels. When we combine the pronunciation assessment service with GPT-4o, the feedback quality is improved, suggesting that the combination of LMMs and traditional models can enhance feedback quality and also provide more accurate scores across both small and large granularities.

\section{Conclusion}

In this study, we examined the potential of LMMs for pronunciation assessment tasks. Our evaluation was primarily centered on the GPT-4o model, using the Speechocean762 dataset, with a specific emphasis on scoring and feedback generation. The findings showcased the zero-shot capabilities of LMMs in pronunciation assessment tasks and provided insights into the model's strengths as well as areas needing further development. While the model is proficient in generating feedback at higher-level granularities, it does not perform as well in scoring at lower-level granularities. However, incorporating the pronunciation assessment service with LMMs can not only generate more diverse and helpful feedback, but also provide accurate scores on smaller granularities for language learners.

\section{Limitations}

The primary limitation of this study is the lack of ground-truth feedback for the speech clips. However, it's important to note that securing reliable human-labeled feedback can often be challenging, as providing feedback is highly subjective and can significantly vary depending on the individual's background and the aspects they concentrate on. Furthermore, we did not fine-tune the LMMs with domain-specific datasets, which could potentially limit the models' performance. In our future research, we aim at addressing these limitations by collecting more human-labeled feedback and fine-tuning the LMMs with domain-specific datasets. This approach will assist in enhancing the performance of pronunciation assessment tasks across both low and high-level granularities and allow for a more thorough exploration of the potential of LMMs in pronunciation assessment tasks.

\bibliographystyle{IEEEtran}
\bibliography{mybib}

\appendix
\onecolumn
\setcounter{table}{0}
\setcounter{figure}{0}
\setcounter{section}{0}
\setcounter{equation}{0}
\renewcommand{\thetable}{A\arabic{table}}
\renewcommand{\thefigure}{A\arabic{figure}}
\renewcommand{\theequation}{A\arabic{equation}}

\section{Prompts}
\label{app:prompts}

Table \ref{tab:prompts} provides the prompts used for evaluating GPT-4o in pronunciation assessment tasks within this study. These tasks encompass scoring, feedback generation, and feedback assessment.

\begin{table*}[th]
  \caption{Prompts for GPT-4o evaluation on pronunciation assessment tasks.}
  \label{tab:prompts}
  \centering
  \begin{tabular}{p{0.97\linewidth}}
    \toprule
    \textbf{Head:} You are tasked with evaluating English pronunciation. For the given audio, the reference text is \textit{\{Reference Text\}}. \\
    \hline
    \textbf{Head with Phoneme:} \textit{\{Head\}}. The reference phonemes are \textit{\{Referecne Phonemes\}}. The ``\textbar'' in the reference phonemes indicates word boundaries, and the reference phonemes are using the CMU dictionary format. \\
    \midrule
    \textbf{Phone:} \textit{\{Head with Phoneme\}}.You need to provide a score from 0 to 10, including phoneme-level accuracy scores for each phoneme. \\
    Our rubric is as follows: \textit{\{Manual Scoring Metrics of Speechocean762 for phoneme\}} \\
    Let‘s think step by step and please return the results in the following JSON format: \textit{\{JSON Template\}}. \\
    \hline
    \textbf{Word:} \textit{\{Head\}}. You need to provide scores from 0 to 10, including word-level accuracy scores, stress scores for each word, as well as total score from 0 to 10. \\
    Our rubric is as follows: \textit{\{Manual Scoring Metrics of Speechocean762 for word\}} \\
    Let's think step by step and please return the results in the following JSON format: \textit{\{JSON Template\}}\\
    \hline
    \textbf{Sentence:} \textit{\{Head\}}. You need to provide scores from 0 to 10, including evaluating sentence-level accuracy, fluency, prosody, and completeness, and provide overall total score from 0 to 10. \\
    Our rubric is as follows: \textit{\{Manual Scoring Metrics of Speechocean762 for sentence\}} \\
    Let's think step by step and please return the results in the following JSON format: \textit{\{JSON Template\}}\\
    \hline
    \textbf{Multigranularity:} \textit{\{Head with Phoneme\}}. You need to provide scores from 0 to 10, including phoneme-level accuracy scores for each phoneme, word-level accuracy scores, and stress scores for each word, as well as total score for 0 to 10. Additionally, you need to evaluate sentence-level accuracy, fluency, prosody, and completeness, and provide overall total score from 0 to 10. Please also provide feedback on the audio according to the scores. \\
    Our rubric is as follows: \textit{\{Manual Scoring Metrics of Speechocean762\}} \\
    Let's think step by step and return the results in the following JSON format: \textit{\{JSON Template\}}. \\
    \hline
    \textbf{Feedback Generation:} \textit{\{Head\}}. You need to provide feedback of the giving audio. \\
    Let's think step by step and return the results in the following JSON format: \textit{\{JSON Template\}}. \\
    \midrule
    \textbf{Feedback Generation with Azure PA:} \textit{\{Head\}}. For the sentence-level score, accuracy is *, fluency is *, prosody is *, completeness is * and the total score is *. For word '*', the accuracy score is *. For word '*', the accuracy score is * ... You need to provide feedback of the giving audio. \\
    Let's think step by step and return the results in the following JSON format: \textit{\{JSON Template\}}. \\
    \midrule
    \textbf{Feedback Assessment:} As an English teacher, you are to evaluate the helpfulness and correlation of feedback with human scores on phoneme-level accuracy, word-level accuracy, stress and total score, and sentence-level accuracy, fluency, prosody, completeness and total score. Helpful feedback should explain what the errors are, why they are errors, and how to fix them. Give a score between 0 and 10, where 0 means the feedback is not helpful at all, and 10 means the feedback is extremely helpful. Similarly, provide a score for the correlation. \\
    \# Content: \textit{\{Reference Text\}} \\
    \# Feedback: \textit{\{Generated Feedback\}} \\
    \# Human Score: \textit{\{Human Scores on Phoneme-, Word- and Sentence-level\}} \\
    Provide output in the following format without explanation: \{``helpfulness'': *, ``correlation'': *\} \\
    \bottomrule
  \end{tabular}
\end{table*}

\newpage
\section{Feedback Assessment}
\label{app:feedback}

In this section, we showcase the sample code employed for feedback assessment utilizing the Azure service. The sample code for a single instance of feedback is meticulously detailed in Algorithm \ref{app:feedback_algo}. The code is designed to appraise the helpfulness and correlation of feedback generated by the GPT-4o for a given audio clip. The feedback is critically evaluated based on human-labeled scores at the phoneme, word, and sentence levels. This code snippet exemplifies the process of assessing the feedback generated by GPT-4o, and accordingly provides a score for both the helpfulness and correlation of the feedback.

\begin{algorithm}
  \caption{Feedback Assessment with Azure service.}
  \label{app:feedback_algo}
  \begin{algorithmic}[1]
    \STATE client = ChatCompletionsClient( \\
    \STATE \hspace{5mm} endpoint=endpoint, \\
    \STATE \hspace{5mm} credential=AzureKeyCredential(credential), \\
    \STATE ) \\
    \STATE response = client.complete( \\
    \STATE \hspace{5mm} messages = [ \\
    \STATE \hspace{10mm} SystemMessage( \\
    \STATE \hspace{15mm} content=( \\
    \STATE \hspace{20mm} ``As an English teacher, you are to evaluate the helpfulness and correlation of feedback with " \\
    \STATE \hspace{20mm} ``human scores on phoneme-level accuracy, word-level accuracy, stress and total score, and " \\
    \STATE \hspace{20mm} ``sentence-level accuracy, fluency, prosody, completeness and total score. " \\
    \STATE \hspace{20mm} ``Helpful feedback should explain what the errors are, why they are errors, and how to fix them. " \\
    \STATE \hspace{20mm} ``Give a score between 0 and 10, where 0 means the feedback is not helpful at all, and 10 means " \\
    \STATE \hspace{20mm} ``the feedback is extremely helpful. Similarly, provide a score for the correlation." \\
    \STATE \hspace{15mm} ) \\
    \STATE \hspace{10mm} ), \\
    \STATE \hspace{10mm} UserMessage( \\
    \STATE \hspace{15mm} content=( \\
    \STATE \hspace{20mm} ``\# Content: however he is not legally responsible" \\
    \STATE \hspace{20mm} ``\# Feedback: The pronunciation showed some inaccuracies, especially with the words `legally' and " \\
    \STATE \hspace{20mm} ```responsible'. The stress and intonation were somewhat off in places, but overall, the speech was understandable." \\
    \STATE \hspace{20mm} ``\# Human Score: For the sentence-level score, accuracy is 7, fluency is 7, prosody is 6, " \\
    \STATE \hspace{20mm} ``completeness is 10 and the total score is 6. " \\
    \STATE \hspace{20mm} ``For word 'however', all word-level scores are 10. " \\
    \STATE \hspace{20mm} ``The phonemes are 'HH AW0 EH1 V ER0'. The accuracy score for the second phoneme is 9, " \\
    \STATE \hspace{20mm} ``while the accuracy scores for the other phonemes are 10. " \\
    \STATE \hspace{20mm} ``For word 'he', all word-level and phoneme-level scores are 10. " \\
    \STATE \hspace{20mm} ``For word 'is', all word-level scores are 10. The phonemes are 'IH1 Z'. The accuracy score for " \\
    \STATE \hspace{20mm} ``the second phoneme is 9, while the accuracy score for the first phoneme is 10. " \\
    \STATE \hspace{20mm} ``For word 'not', all word-level and phoneme-level scores are 10. " \\
    \STATE \hspace{20mm} ``For word 'legally', the accuracy score is 5, stress score is 10, and the total score score is 6. " \\
    \STATE \hspace{20mm} ``The phonemes are 'L IY1 G AH0 L IY0'. The accuracy scores for the third phoneme is 6, the accuracy " \\
    \STATE \hspace{20mm} ``score for the fourth phonemes is 6, the accuracy score for the fifth phoneme is 4, " \\
    \STATE \hspace{20mm} ``and the accuracy scores for the other phonemes are 10. " \\
    \STATE \hspace{20mm} ``For word 'responsible', all word-level scores are 10. " \\
    \STATE \hspace{20mm} ``The phonemes are 'R IH0 S P AH1 N S AH0 B L'. The accuracy score for the second phoneme is 6, " \\
    \STATE \hspace{20mm} ``while the accuracy scores for the other phonemes are 10." \\
    \STATE \hspace{20mm} ``Provide output in the following format without explanation: \{'helpfulness': *, 'correlation': *\}" \\
    \STATE \hspace{15mm} ) \\
    \STATE \hspace{10mm} ), \\
    \STATE \hspace{5mm} ], \\
    \STATE \hspace{5mm} temperature=0, \\
    \STATE \hspace{5mm} top\_p=1, \\
    \STATE \hspace{5mm} max\_tokens=15, \\
    \STATE )
    \STATE print(``Response:", response.choices[0].message.content.strip())
  \end{algorithmic}
\end{algorithm}

\end{document}